\documentclass[aps,prb,twocolumn,superscriptaddress,showpacs,showkeys]{revtex4}
\usepackage{bm,amsmath,graphicx,dcolumn}
\begin{document}

\title{Strain field of InAs QDs on GaAs (001) substrate surface: characterization by synchrotron X-ray Renninger scanning}
\author{S.L. Morelh\~ao}
\email{morelhao@if.usp.br}
\author{L.H. Avanci}
\author{R. Freitas}
\author{A.A. Quivy}
\affiliation{Instituto de F\'{\i}sica, Universidade de S\~ao Paulo, CP 66318, 05315-970 S\~aoPaulo, SP, Brazil}

\date{\today}
\begin{abstract}
Precise lattice parameter measurements in single crystals are achievable, in principle, by X-ray multiple diffraction (MD) experiments. Tiny sample misalignments can compromise systematic usage of MD in studies where accuracy is an important issue. In this work, theoretical treatment and experimental methods for correcting residual misalignment errors are presented and applied to probe the induced strain of buried InAs quantum dots on GaAs (001) substrates.
\end{abstract}



\maketitle

\section{Introduction}

Opto-electronic devices based on self-organized InAs quantum dots (QDs) are very suitable for metropolitan-area network applications. Their wavelength emission range match the absorption minima of the optical fibers; this optical property is still under improvement to provide an even better emission-absorption match.~\cite{silva2003} The actual challenge resides in increasing the optical efficiency of these devices, which is possible by increasing the density of optically active QDs. Recent studies have demonstrated drastic reduction in the number of optically active structures after the growth of a few atomic GaAs layers over the QDs. Since surface probe techniques are no longer useful to inspect the physical structure of buried QDs; alternative structural characterization procedures have become relevant.

X-ray Renninger scanning~\cite{renn1937} (XRS) is the most precise technique for accurate lattice parameter determination in single crystals.~\cite{post1975} However, in practice, even in very good equipments and careful alignment procedures, some tiny sample misalignments are always present. They have compromised XRS application in studies where accuracy is an important issue. In this work, we describe how to consistently correct systematic errors in XRS to achieved high sensitivity in the lattice-parameter variation ($\Delta a/a\simeq10^{-5}$). Combining such an accuracy with the shallow penetration depth of the X-ray wavefield under Bragg-surface diffraction~\cite{more1999} \textemdash~a particular multiple wave diffraction case~\textemdash~ the residual average strain field on the substrate lattice of devices with buried QDs has been characterized along two orthogonal in-plane directions.

\section{X-ray diffraction geometry}

X-ray multiple diffraction (MD) in crystals are excited when the incident beam direction, wavevector $\bm{k}$, fulfill two conditions summarized by

\begin{figure}
\includegraphics[width=3.2in]{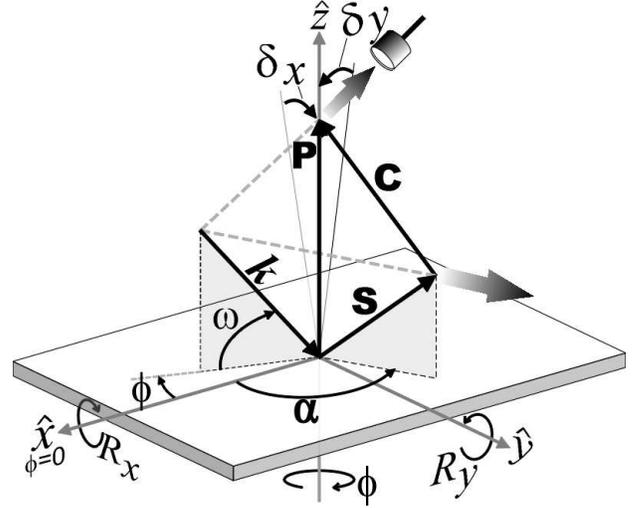}
\caption{\label{fig1} Three-beam X-ray diffraction in crystals with the primary diffraction vector \textbf{P} aligned along the scanning axis $\phi$, $\textbf{P}\parallel\hat{z}$. When a secondary reflection S diffracts, $\textbf{C}=\textbf{P}-\textbf{S}$ provides the energy coupling $\text{S}\Rightarrow\text{P}$ reflection. Rotation matrices $R_x(\delta_x)$ and $R_y(\delta_y)$, account for tiny misalignments of diffraction vectors with respect to the $(\hat{x},\>\hat{y},\>\hat{z})$ goniometer's reference system. $\bm{k}$ is the incident beam wavevector.}
\end{figure}

\begin{equation}
\bm{k}\cdot\textbf{P}=-\textbf{P}\cdot\textbf{P}/2
\label{eq1}
\end{equation}
and
\begin{equation}
\bm{k}\cdot\textbf{S}=-\textbf{S}\cdot\textbf{S}/2.
\label{eq2}
\end{equation}
{\bf P} and {\bf S} are the diffraction vectors of the primary and secondary reflections, respectively. In XRS, the primary reflection is kept excited, i.e. Eq.~(\ref{eq1}) fulfilled, while the crystal rotates around {\bf P}. It requires {\bf P} parallel to the goniometer $\phi$ axis,~\cite{more2003a} the $\hat{z}$ axis in Fig. 1. MD occurs when secondary reflections are excited by the $\phi$ rotation and, therefore, Eq.~(\ref{eq2}) is also fulfilled. The primary intensity changes as secondary reflections diffract, e.g. Fig.~2.

The most well known expressions for determining MD positions in XRS were obtained from Eq.~(\ref{eq2})~\cite{cole1962, caticha1975} assuming that the primary reflection is always aligned, i.e. Eq.~(\ref{eq1}) fulfilled during a complete $\phi$ rotation of $360^o$. An alternative approach to determine such positions without the $\textbf{P}\parallel\hat{\bm{z}}$ constrain has been obtained~\cite{more2003b} by writing $\textbf{P}=\textbf{S}+\textbf{C}$ in Eq.~(\ref{eq1}), which leads to

\begin{equation}
\bm{k}\cdot\textbf{C}=-\textbf{C}\cdot\textbf{C}/2-\textbf{C}\cdot\textbf{S}. 
\label{eq3}
\end{equation}

Sample misalignments $\delta_x$ and $\delta_y$, as shown in Fig.~1, generate small $\Delta\omega=\omega-\omega_0$ and $\Delta\phi=\phi-\phi_0$ corrections in the incident beam direction $\bm{k}(\omega,\phi)\simeq\bm{k}_0+\Delta\omega\bm{k}_{\omega}+\Delta\phi\bm{k}_{\phi}$. These corrections can be determined by a trivial system of linear equations

\begin{equation}
\left[\begin{array}{cc}
\bm{k}_{\omega}\cdot\textbf{S} & \bm{k}_{\phi}\cdot\textbf{S}\\
\bm{k}_{\omega}\cdot\textbf{C} & \bm{k}_{\phi}\cdot\textbf{C}\end{array}\right]
\left[\begin{array}{c}
\Delta\omega\\
\Delta\phi\end{array}\right]=-
\left[\begin{array}{c}
(\textbf{S}/2+\bm{k}_0)\cdot\textbf{S}\\
(\textbf{C}/2+\textbf{S}+\bm{k}_0)\cdot\textbf{C}\end{array}\right]
\label{eq4}
\end{equation}
derived from Eqs.~(\ref{eq2}) and (\ref{eq3}). $\bm{k}_{\omega}=\partial\bm{k}/\partial\omega$ and $\bm{k}_{\phi}=\partial\bm{k}/\partial\phi$ are calculated at $\omega_0$ and $\phi_0$, the incidence and azimuthal angles for exciting the MD in non-misaligned samples, i.e. at $\bm{k}_0=\bm{k}(\omega_0,\phi_0)$. Moreover, $\textbf{G}=\textbf{G}_0+\Delta\textbf{G}$ stands for the $\textbf{S}$ or $\textbf{C}$ diffraction vectors where \[\Delta\textbf{G}=({\rm G}_z\delta_y,~{\rm G}_z\delta_x,~-{\rm G}_x\delta_y-{\rm G}_y\delta_x)\] and $\textbf{G}_0=({\rm G}_x,~{\rm G}_y,~{\rm G}_z)$.

In non-misaligned sample the MD positions are obtained from Eq.~(\ref{eq2}) alone, which
provides

\begin{equation}
\cos(\phi_0-\alpha)=\cos\beta=\frac{\lambda|\textbf{S}|/2-{\rm S}_z\sin\omega_0}{{\rm S}_{xy}\cos\omega_0}
\label{eq5}
\end{equation}
where $\textbf{S}_x=\textbf{S}_{xy}\cos\alpha=\hat{x}\cdot\textbf{S}_0$, $\textbf{S}_y=\textbf{S}_{xy}\sin\alpha=\hat{y}\cdot\textbf{S}_0$, and $\omega_0$ is the Bragg angle of reflection P. 

There are two azimuthal positions where the same secondary reflection is excited: $\phi_1=\alpha-\beta$ and $\phi_2=\alpha+\beta$. Lattice-parameter determination is based on the experimental measurements of both positions. They provide $2\beta_{exp}=\phi_{2,exp}-\phi_{1,exp}$ for the calculation of the unit cell parameters from Eq.~(\ref{eq5}), as described elsewhere (see for instance Ref.~\onlinecite{baker1975}). The residual sample misalignments are measured by rocking curves of the primary reflection at $\phi=0$, $90^o$, $180^o$, and $270^o$. If $\omega_0$, $\omega_{90}$, $\omega_{180}$ and $\omega_{270}$ are the respective rocking-curve peak positions after the final alignment,

\begin{equation}
\delta_x=(\omega_{270}-\omega_{90})/2\textrm{~~~and~~~}
\delta_y=(\omega_{180}-\omega_0)/2.
\label{eq10}
\end{equation}

The $\Delta\phi_{n=1,2}$ corrections in the azimuthal positions are obtained from Eq.~(\ref{eq4}), and they can be used either to refine the $2\beta_{exp}$ value according to

\begin{equation}
2\beta_{exp}=(\phi_{2,exp}-\Delta\phi_2)-(\phi_{1,exp}-\Delta\phi_1)
\label{eq11}
\end{equation}
or to estimate misalignment effects on different secondary reflections, for instance those providing some useful information about the crystalline structure of the sample. 

\section{Experimental}

Data collection has been carried out at the Brazilian Synchrotron Light Laboratory (LNLS) D12A (XRD-1) beam-line, with the polarimeter-like diffractomer;~\cite{more2003a} smallest step size is $0.0004^o$ in both $\omega$ and $\phi$ axes. The wavelength, $\lambda=1.330234$\AA ~from a Si (111) double-crystal monochromator, was measured by rocking-curves of the 111 and 333 silicon reflections. Vertical incidence plane scattering upwards ($\sigma$-polarization). The sample is a commercial GaAs (001) substrate with InAs quantum dots (QD) grown by molecular beam epitaxy (MBE) using a rate of 0.009 monolayer per second (mL/s). There is a $300$\AA ~thick GaAs cap layer.

\begin{figure} 
\includegraphics[width=3.2in]{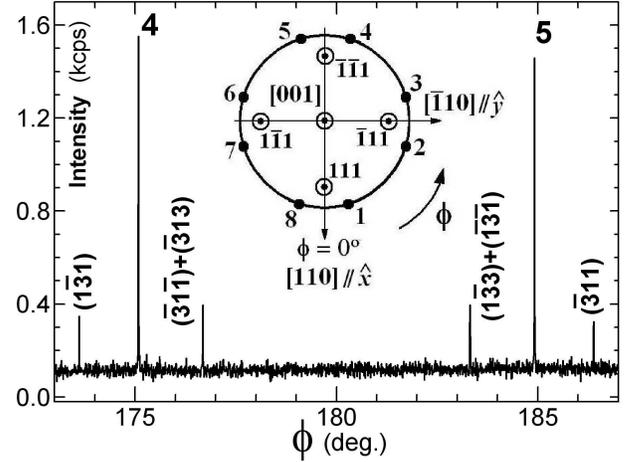}
\caption{\label{fig2} XRS of the 002 GaAs reflection carried out with X-ray wavelength $\lambda=1.330234$\AA. The peaks at $\phi=175.08^o$ (No. 4) and $184.92^o$ (No. 5) occur when the $\bar{1}11$ and $1\bar{1}1$ secondary reflections are excited, respectively. MD positions (closed circles, 1 to 8) owing to other secondary reflections of the {111} family are indicated in the inset; their measured $\phi$ positions are provided in Table 1.}
\end{figure}

\section{Results and discussions}

Table~\ref{tab1} shows the $\phi_1$ and $\phi_2$ positions of the $111$, $\bar{1}11$, $\bar{1}\bar{1}1$, and $1\bar{1}1$ secondary reflections in the XRS of the 002 GaAs reflection. The peak positions were determined by fitting the intensity data with lorentzian-gaussian convolution curves: a lorentzian function standing for the intrinsic profile (FWHM $\simeq0.0048^o$) while a gaussian accounts for the instrumental broadening (FWHM $\simeq0.0060^o$). A short portion of the XRS around $\phi=180^o$ is shown in Fig. 2; the $[110]$ direction is the reference for $\phi=0$. To check for instrumental errors, $\phi$-scans of the peaks 4 and 5 were repeated several times after $\pm 360^o$ rotations in $\phi$; the observed differences were not larger than $0.00075^o$. Each $\phi$ scan has been performed at the maximum of the (002) rocking-curve, which was carried out about $0.5^o$ before the MD peak positions.

The rocking curves used to characterize the residual misalignments are shown in Fig. 3; they provide $\delta_x\simeq0.0009^o$ and $\delta_y\simeq0.0067^o$. Since these values are very small, Eq.~(\ref{eq4}) can be linearized by numerical derivation, and the azimuthal corrections written as $\Delta\phi_n={\rm A}_n\delta_x+{\rm B}_n\delta_y$. The 4-fold axis symmetry of the measured secondary reflections establish some relationships among the ${\rm A}_n$ and ${\rm B}_n$ coefficients, as given in Table~\ref{tab1} (last two rows). By replacing them into Eq.~(\ref{eq11}), it is possible to demonstrate that

\begin{subequations}
\begin{equation}
\bar{\beta}_{[110]}=(\beta_{111}+\beta_{\bar{1}\bar{1}1})/2
\label{eq12a}
\end{equation}
and 
\begin{equation}
\bar{\beta}_{[\bar{1}10]}=(\beta_{\bar{1}11}+\beta_{1\bar{1}1})/2
\label{eq12b}
\end{equation}
\label{eq12}
\end{subequations}
do not depend on $\delta_x$ and $\delta_y$. In other words, the average $\beta_{exp}$ for the secondary reflections with in-plane components along the $[110]$ and $[\bar{1}10]$ orthogonal directions, $\bar{\beta}_{[110]}$ and $\bar{\beta}_{[\bar{1}10]}$, are misalignment-free experimental values and, therefore, useful for precise lattice parameter measurements in both directions.

\begin{figure} 
\includegraphics[width=3.2in]{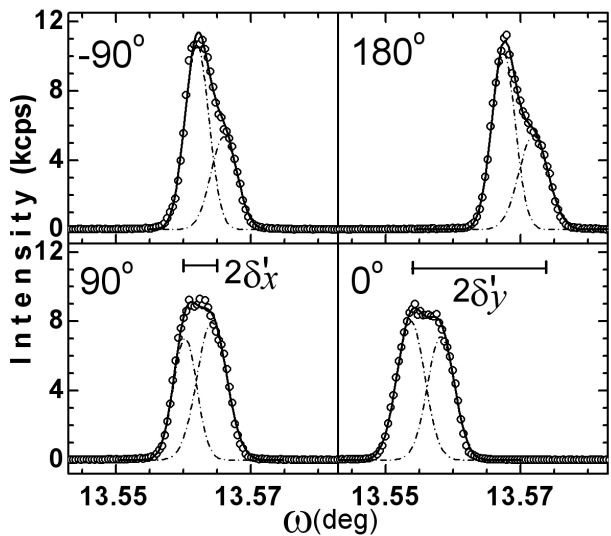}
\caption{\label{fig3} Experimental (open circles) and simulated (solid lines) rocking curves, $\omega$-scans, of the 002 primary reflection measured at $\phi=0$, $90^o$, $180^o$ and $270^o$ (top-left of each scan). The scans were simulated by two gaussian functions (dashed lines) whose splitting is about $\Delta\omega\simeq-0.00422(24)^o$, and hence $\Delta c/c \simeq 3.0(2) \times 10^{-4}$. $\delta_x^{\prime}~(=0.0024^o)$ and $\delta_y^{\prime}~(=0.0103^o)$ are the expected misorientations according to the $\bar{\beta}$ values in Table~\ref{tab1}.}
\end{figure}

Assuming a tetragonal substrate lattice distortion, its in-plane lattice parameter is given by $a=b=(1-\nu)a_0$ while, to preserve the unit cell volume, $c=(1+2\nu)a_0$. For $a_0=5.6534$\AA, $|\textbf{P}|=2/c$, $\textbf{S}_0=(h/a,~k/b,~1/c)$, and $h,k=\pm1$, Eq.~(\ref{eq5}) provides

\begin{equation}
\beta(\nu)=85.08984^o-33.85^o\nu-5.2^o\tfrac{\Delta\lambda}{\lambda}.
\label{eq13}
\end{equation}

Note that $|\textbf{S}|$=$|\textbf{S}_0|$ is invariant under any rotation, but their components are not. Therefore, ${\rm S}_x$, ${\rm S}_y$ and ${\rm S}_z$ in Eq.~(\ref{eq5}) must be calculated for $\hat{x}=(1,~1,~0)/\sqrt{2}$, $\hat{y}=(\bar{1},~1,~0)/\sqrt{2}$ and $\hat{z}=(0,~0,~1)$, \textit{i.e.}, ${\rm S}_x=(h+k)/(a\sqrt{2})$, ${\rm S}_y=(-h+k)/(a\sqrt{2})$ and ${\rm S}_z=1/c$.

\begin{table*} 
\caption{\label{tab1}Azimuthal $\phi$ positions of the $111$, $\bar{1}\bar{1}1$, $\bar{1}11$, and $1\bar{1}1$ secondary reflections in the XRS of the 002 GaAs reflection. Each position was measured three times (rows 1, 2 and 3) as explained in the text, $\bar{\phi}=(\phi_{max}+\phi_{min})/2$, $\varepsilon=(\phi_{max}-\phi_{min})/2$, $\beta_{exp}=(\bar{\phi}_2-\bar{\phi}_1)/2$, $\bar{\beta}$ are the average misalignment-free values, $\nu$ stands for the unit-cell tetragonal distortion [Eq.~(\ref{eq13})] and $\beta_{mis}=\bar{\beta}+(\Delta\phi_2-\Delta\phi_1)/2$ [Eq.~(\ref{eq11})] where $\Delta\phi_n={\rm A}_n\delta_x^{\prime}+{\rm B}_n\delta_y^{\prime}$, $\delta_x^{\prime}=0.0024^o$, and $\delta_y^{\prime}=0.0103^o$. ${\rm A}_n$ and ${\rm B}_n$ were estimated by numerical derivation of Eq.~(\ref{eq4}). Angular values are given in degrees.}
\begin{ruledtabular} 
\begin{tabular}{ccccccccc}
&\multicolumn{2}{c}{$111$}&\multicolumn{2}{c}{$\bar{1}\bar{1}1$}&\multicolumn{2}{c}{$\bar{1}11$}&\multicolumn{2}{c}{$1\bar{1}1$}\\
 & $\phi_1$ & $\phi_2$ & $\phi_1$ & $\phi_2$ & $\phi_1$ & $\phi_2$ & $\phi_1$ & $\phi_2$\\ \hline
1&-85.07750&85.08599&94.91867&265.09121&4.92085&175.08692&184.92072&355.08935\\ 
2&-85.07716&85.08575&94.91853&265.09141&4.92102&175.08688&184.92061&355.08881\\
3&-85.07683&85.08548&94.91819&265.09209&4.92105&175.08730&184.92064&355.08880\\
$\bar{\phi}$&-85.077165&85.085735&94.91843&265.09165&4.92095&175.08709&184.920665&355.089075\\
$\varepsilon$&$\pm$0.000335&$\pm$0.000255&$\pm$0.00024&$\pm$0.00044&$\pm$0.00010&$\pm$0.00021&$\pm$0.000055&$\pm$0.000275\\
$\beta_{exp}$&\multicolumn{2}{c}{85.081450$\pm$0.000295}&\multicolumn{2}{c}{85.08661$\pm$0.00034}&\multicolumn{2}{c}{85.083070$\pm$0.000155}&\multicolumn{2}{c}{85.084205$\pm$0.000165}\\
$\bar{\beta}$&\multicolumn{4}{c}{85.08403$\pm$0.00032}&\multicolumn{4}{c}{85.08364$\pm$0.00016}\\
$\nu$&\multicolumn{4}{c}{$(1.716\pm0.096)\times 10^{-4}$}&\multicolumn{4}{c}{$(1.832\pm0.048)\times 10^{-4}$}\\
$\beta_{mis}$&\multicolumn{2}{c}{85.0815}&\multicolumn{2}{c}{85.0865}&\multicolumn{2}{c}{85.0831}&\multicolumn{2}{c}{85.0842}\\
${\rm A}_n$&0.0207&0.0207&-0.0207&-0.0207&0.2412&-0.2412&-0.2413&0.2413\\
${\rm B}_n$&0.2412&-0.2412&-0.2413&0.2413&-0.0207&-0.0207&0.0207&0.0207\\
\end{tabular}
\end{ruledtabular}
\end{table*}

\begin{figure} 
\includegraphics[width=3.2in]{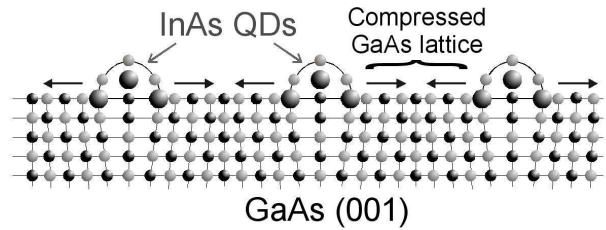}
\caption{\label{fig4} Schematic illustration of the compressive stress generated by InAs QDs on GaAs substrate. There is also a 300\AA ~thick GaAs cap layer over the QDs (not shown in the figure).} 
\end{figure}

The observed tetragonal deformation $\nu\simeq1.77\times 10^{-4}$, corresponds to an average value in a shallow layer just below the surface, not thicker than $0.3\>\mu m$,~\cite{more1999} where the induced strain due to the InAs QD is significant. Although the InAs epitaxial growth generates an expansive stress in the substrate lattice under the QDs, the adjacent regions are compressed as schematically illustrated in Fig.~\ref{fig4}. This in-plane compressive strain may have been propagated into the epitaxial cap layer.

An error in the wavelength value of $\Delta\lambda=+0.0015\text{\AA}$ could be responsible for the $\bar{\beta}$ values obtained in Table 1; but, the used method for determining $\lambda$ assures a precision that is at least 10 times better, $\Delta\lambda/\lambda < 2.0\times 10^{-4}$.

Differences in the $\nu$ values along the $[110]$ and $[\bar{1}10]$ directions, \textit{i.e.} $\bar{\beta}_{[110]}\neq\bar{\beta}_{[\bar{1}10]}$, would yield a unit cell slightly twisted near the substrate surface. The variation in the $\gamma$ angle, between the $\bm{a}$ and $\bm{b}$ lattice vectors, is estimated from Eq.~(\ref{eq5}) as 

\begin{equation}
\bar{\beta}_{[\bar{1}10]}-\bar{\beta}_{[110]}=(C_{\bar{1}10}-C_{110})\Delta\gamma
\label{eq14}
\end{equation}
where $C_{\bar{1}10}=0.1276$ and $C_{110}=-0.1302$; then $\gamma=89.9985^o\pm0.0018^o$. Non-uniform self-organization of QDs plus the movement of the beam spot on the sample surface during the XRS could produce similar results. It could also be responsible by the observed disagreement between the $\delta_{x,y}$ and $\delta_{x,y}^{\prime}$ values.

\section{Conclusions}

In summary, refinement of XRS data for residual sample misalignment errors has led to a precise tool for probing in-plane lattice strains, as small as $\Delta a / a = 10^{-5}$. Although, the strain of InAs QDs on the surface of GaAs (001) substrates was the subject investigated here, the method can be extended for studying other epitaxial nanostructures by their strain field on the substrate lattice.

\begin{acknowledgments}
This work was supported by the Brazilian founding agencies FAPESP (proc. No. 02/10387-5 and 02/10185-3), CNPq (proc. No. 301617/95-3 and 150144/03-2) and LNLS (under proposal No. D12A-XRD1 2490/03). 
\end{acknowledgments}


\begin{thebibliography}{9}.
\bibitem{silva2003} M.J. da Silva et al, Appl. Phys. Lett. {\bf 82}, 2646 (2003).
\bibitem{post1975} B. Post, J. Appl. Cryst. {\bf 8}, 452 (1975).
\bibitem{more1999} S. L. Morelh\~ao and E. Abramof, J. Appl. Cryst. \textbf{33}, 871 (1999).
\bibitem{more2003a} S. L. Morelh\~ao, J. Sync. Rad. \textbf{10}, 236 (2003).
\bibitem{renn1937} M. Renninger, Z. Kristallogr. \textbf{97}, 107 (1937).
\bibitem{cole1962} H. Cole, F. W. Chambers and H. M. Dunn, Acta Cryst. \textbf{15}, 138 (1962).
\bibitem{caticha1975} S. Caticha Ellis, Jpn. J. Appl. Phys. \textbf{14}(5), 603 (1975).
\bibitem{more2003b} S. L. Morelh\~ao, A. A. Quivy and J. H\"artwig, Microel. Journal \textbf{34}, 695 (2003)
\bibitem{baker1975} J. F. C. Baker and M. Hart, Acta Cryst. A\textbf{31}, 364 (1975).
\end{thebibliography}
\end{document}